\documentclass[12pt,twoside]{article}
\usepackage{fleqn,espcrc1}

\usepackage{graphicx}
\usepackage[figuresright]{rotating}

\newcommand{\AmS}{{\protect\the\textfont2
  A\kern-.1667em\lower.5ex\hbox{M}\kern-.125emS}}

\title{Systematic Study of Au-Au Collisions with AGS Experiment E917}
\begin{document}
    \def\thefootnote{\fnsymbol{footnote}}%
% typeset front matter
\maketitle
\noindent
 B.~Holzman$^{\rm a}$, for the E917 Collaboration\footnote{This 
 work was supported by the U.~S. Department of Energy, 
 the National Science Foundation (USA) and KOSEF (Korea).} \\

\noindent %
B.B.~Back$^{\rm b}$, R.R.~Betts$^{\rm a}$, J.~Chang$^{\rm c}$, 
W.C.~Chang$^{\rm c}$,
C.Y.~Chi$^{\rm d}$, Y.Y.~Chu$^{\rm e}$, J.B.~Cumming$^{\rm e}$, 
J.C.~Dunlop$^{\rm f}$, 
W.~Eldredge$^{\rm c}$, S.Y.~Fung$^{\rm c}$, R.~Ganz$^{\rm a}$, 
E.~Garcia$^{\rm g}$,
A.~Gillitzer$^{\rm b}$, G.~Heintzelman$^{\rm f}$, W.F.~Henning$^{\rm b}$, 
D.J.~Hofman$^{\rm a}$, B.~Holzman$^{\rm a}$,
J.H.~Kang$^{\rm h}$, E.J.~Kim$^{\rm h}$,
S.Y.~Kim$^{\rm h}$, Y.~Kwon$^{\rm h}$, D.~McLeod$^{\rm a}$, 
A.C.~Mignerey$^{\rm g}$, 
M.~Moulson$^{\rm d}$, V.~Nanal$^{\rm b}$, C.A.~Ogilvie$^{\rm f}$, 
R.~Pak$^{\rm i}$,
A.~Ruangma$^{\rm g}$, D.~Russ$^{\rm g}$, R.K.~Seto$^{\rm c}$, 
P.J.~Stanskas$^{\rm g}$,
G.S.F.~Stephans$^{\rm f}$,
H.Q.~Wang$^{\rm c}$, F.L.H.~Wolfs$^{\rm i}$, A.H.~Wuosmaa$^{\rm b}$, 
H.~Xiang$^{\rm c}$,
G.H.~Xu$^{\rm c}$, H.B.~Yao$^{\rm f}$, C.M.~Zou$^{\rm c}$

\noindent
\\~
{}$^{\rm a}$ University of Illinois at Chicago, Chicago, IL 60607 \\
{}$^{\rm b}$ Argonne National Laboratory, Argonne, IL 60439 \\
{}$^{\rm c}$ University of California, Riverside, CA 92521 \\
{}$^{\rm d}$ Nevis Laboratories, Columbia University, Irvington, NY 10533 \\
{}$^{\rm e}$ Brookhaven National Laboratory, Upton, NY 11973 \\
{}$^{\rm f}$ Massachusetts Institute of Technology, Cambridge, MA 02139 \\
{}$^{\rm g}$ University of Maryland, College Park, MD 20742 \\
{}$^{\rm h}$ Yonsei University, Seoul 120-749, South Korea \\
{}$^{\rm i}$ University of Rochester, Rochester, NY 14627 

\begin{abstract}
The systematics of baryon stopping and strange particle production 
have been studied in Au+Au collisions
at 6, 8, and 10.8 GeV per nucleon with AGS experiment E917.
In particular, the systematic behavior of the proton yields, 
$\phi$ meson production, and the yields of antilambdas and antiprotons are
examined as functions of beam energy and centrality.
\end{abstract}

\section{INTRODUCTION}
In just the last year, center-of-mass energies reached in heavy ion 
collisions have increased by more than an order of magnitude over 
those previously attained at the AGS.  Nevertheless, several 
outstanding issues remain
in the AGS energy regime.  The characterization of the evolution of the 
initial-state nucleons can be studied through the spectra and 
rapidity distribution of protons.
The mechanisms underlying the production of strange and anti-strange quarks
may be investigated through the examination of the yields and spectral shapes
of $\phi$ mesons, antilambdas, and antiprotons.

\section{EXPERIMENT}
 E917 is the last in the E802/E859/E866 series of heavy ion experiments 
 that were performed at the AGS.  The E917 apparatus
 consisted of a series of beam
 counters, a beam vertexing detector, a multiplicity array surrounding 
 the target, a hodoscope,
 a zero-degree calorimeter, and a movable spectrometer composed
 of drift chambers, magnet, and time-of-flight wall.  
 The calorimeter and multiplicity array were used to determine
 collision centrality, while the spectrometer was used for tracking and
 particle identification.
 Detailed descriptions of the equipment can be found 
in~\cite{Ahle98,Abbott90}.
\section{PROTON YIELDS}
\begin{figure}[h]
\begin{center}
\includegraphics[width=4.2in, angle=270]{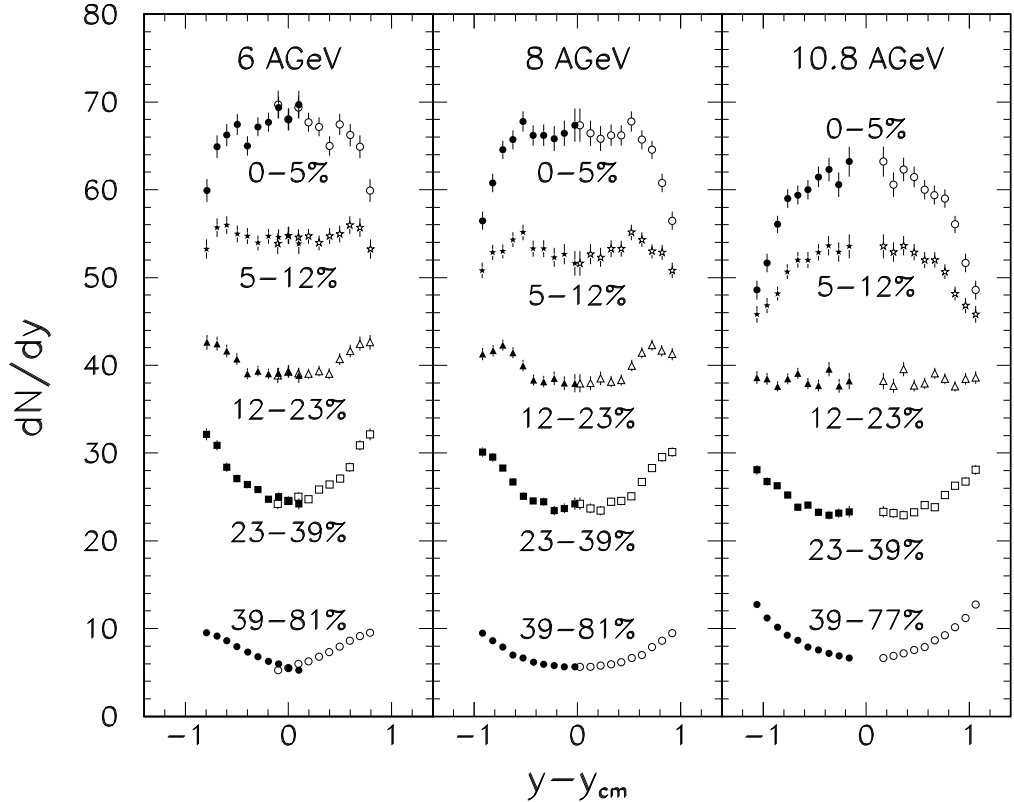}
\caption{Proton rapidity distributions at three beam energies for
five centrality classes.  Open points are the measured data reflected about
mid-rapidity.}
\label{proton_dndy}
\end{center}
\end{figure}

The invariant yields of protons were determined from 6, 8, and 10.8 GeV/nucleon
Au+Au collisions as a function of transverse mass 
($m_T = \sqrt{m_0^2 + p_T^2}$).  The data were fit with a Boltzmann function 
over ten rapidity intervals.   Rapidity distributions ($dN/dy$) were
extracted for five centrality classes at each beam energy and are presented in
Figure~\ref{proton_dndy}.

Two qualitative features emerge from the data.
First, a smooth evolution in shape from
concave for the most peripheral event class ($39\%-77/81\%$) to convex
for the most central ($0-5\%$) is evident at all beam energies.  Second, the
$dN/dy$ curves are not well-described by a singly-peaked function, but
rather are consistent with a bimodal distribution, even for
the most central events.
The nature of this double-peak has been quantified by fitting the curves
with a Gaussian distribution reflected about mid-rapidity~\cite{Back00}.  The persistence
of bimodal shape for even the most central events
suggests that complete
stopping is not achieved at AGS energies, and that the 
longitudinal rapidity distribution is a result of transparency.

\section{ANTILAMBDA/ANTIPROTON RATIO}
A measurement of  the ratio of antilambdas ($\bar{\Lambda}$) to 
directly produced antiprotons ($\bar{p}_{direct}$) allows us to
indirectly compare the ratio of $\bar{s}$ quarks to light anti-quarks
($\bar{u}$, $\bar{d}$).
Theoretical predictions for this quantity range from $0.9$ from 
thermal models~\cite{CleymansComm} to $1.3$ in the RQMD cascade 
model~\cite{WangComm}.

Invariant mass spectra for accepted $\bar{p}\pi^+$ pairs were constructed for
bins in $m_T$ and $y$ for both central ($0-12\%$) and peripheral
($12-77\%$) collisions at a beam energy of $10.8$ GeV/nucleon for
events which contained at least one $\bar{p}$ in the spectrometer.
An event-mixed background was normalized to the mass region outside
the $\bar{\Lambda}$ peak, defined as lying more than 
6 standard deviations away from the nominal $\bar{\Lambda}$ mass.
The yield was determined by subtracting the number of 
$\bar{p}\pi^+$ pairs which lie within 3 standard deviations of the peak from
those in the background.  
Since a significant number of the observed antiprotons are 
produced by $\bar{\Lambda}$ 
decay, both $m_T$ spectra 
were simultaneously fit with Boltzmann functions, 
as described in~\cite{Back01}.  This fit procedure
yields values for $dN/dy_{\bar\Lambda}$, $\bar{\Lambda}/\bar{p}_{direct}$, 
and the inverse slopes for each.

We find that the $\bar{\Lambda}$/$\bar{p}$ ratio increases
from $0.26^{+.19}_{-.15}{}^{+.5}_{-.4}$ in peripheral events to
$3.6^{+4.7}_{-1.8}{}^{+2.7}_{-1.1}$ in central events.  This is 
consistent with an earlier indirect estimate which combined
different results for the $\bar{p}$ yield from the 
E864 and E878 experiments, under the
assumption that the difference was due to different acceptances
for $\bar{p}$ from $\bar{\Lambda}$ feed-down~\cite{e864}. 

\section{PHI MESON YIELDS}
The $\phi$ meson is the lightest bound state
of two strange quarks ($s\bar{s}$); it is therefore a useful probe of
the strangeness produced in the reaction. It's properties may 
also be sensitive to in-medium modifications of meson properties.   

\begin{figure}[h]
\begin{minipage}[b]{.46\linewidth}
 \centering
 \includegraphics[width=\linewidth,clip=true]
{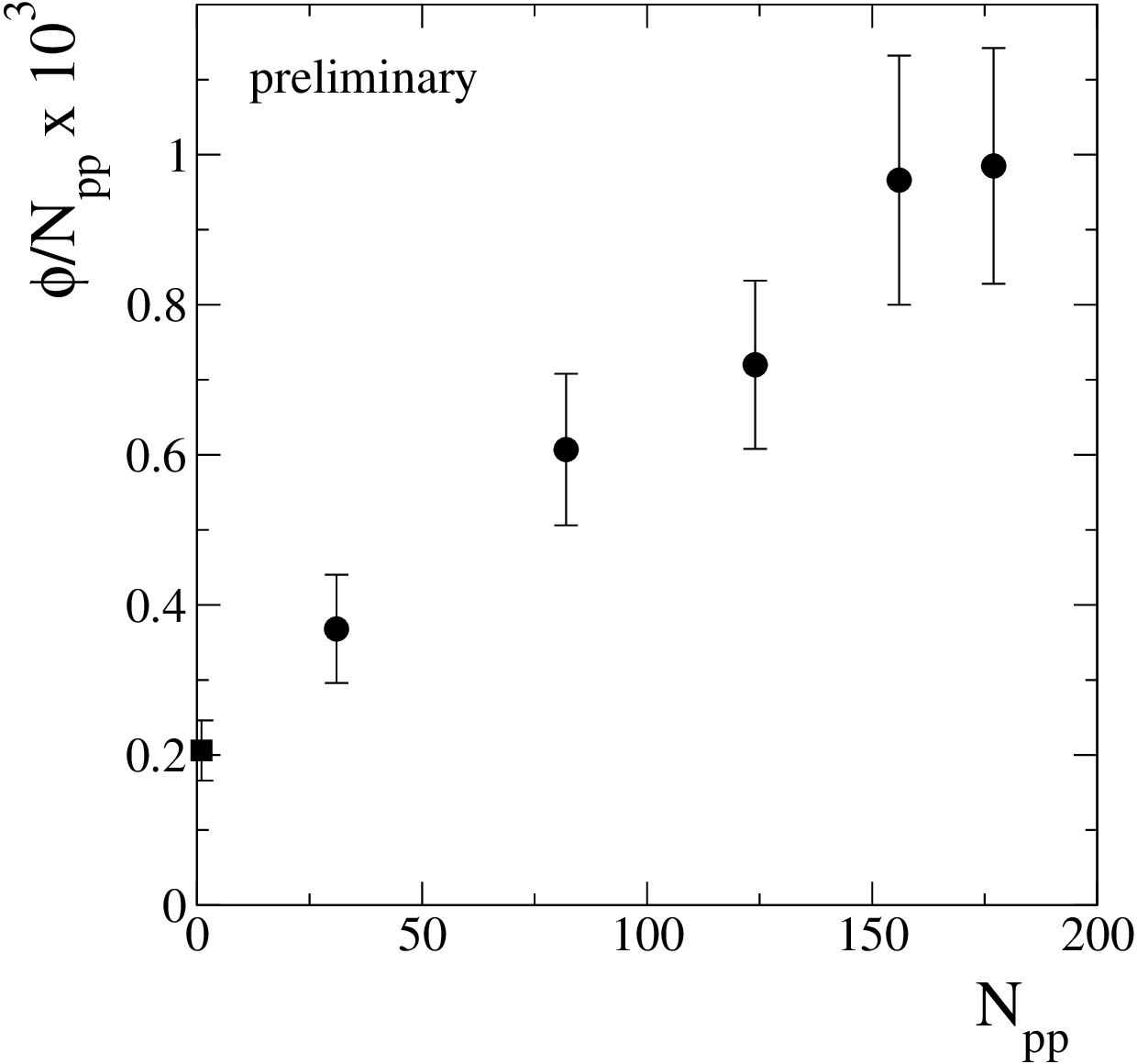}
 \caption{Ratio of $\phi$ yield to $N_{pp}$ as a function of $N_{pp}$.  The
circles are E917 data; the square is p+p from Ref. \cite{phi_pp}.}
 \label{phinpp}
\end{minipage}\hfill
\begin{minipage}[b]{.46\linewidth}
 \centering
 \includegraphics[width=\linewidth,clip=true]
{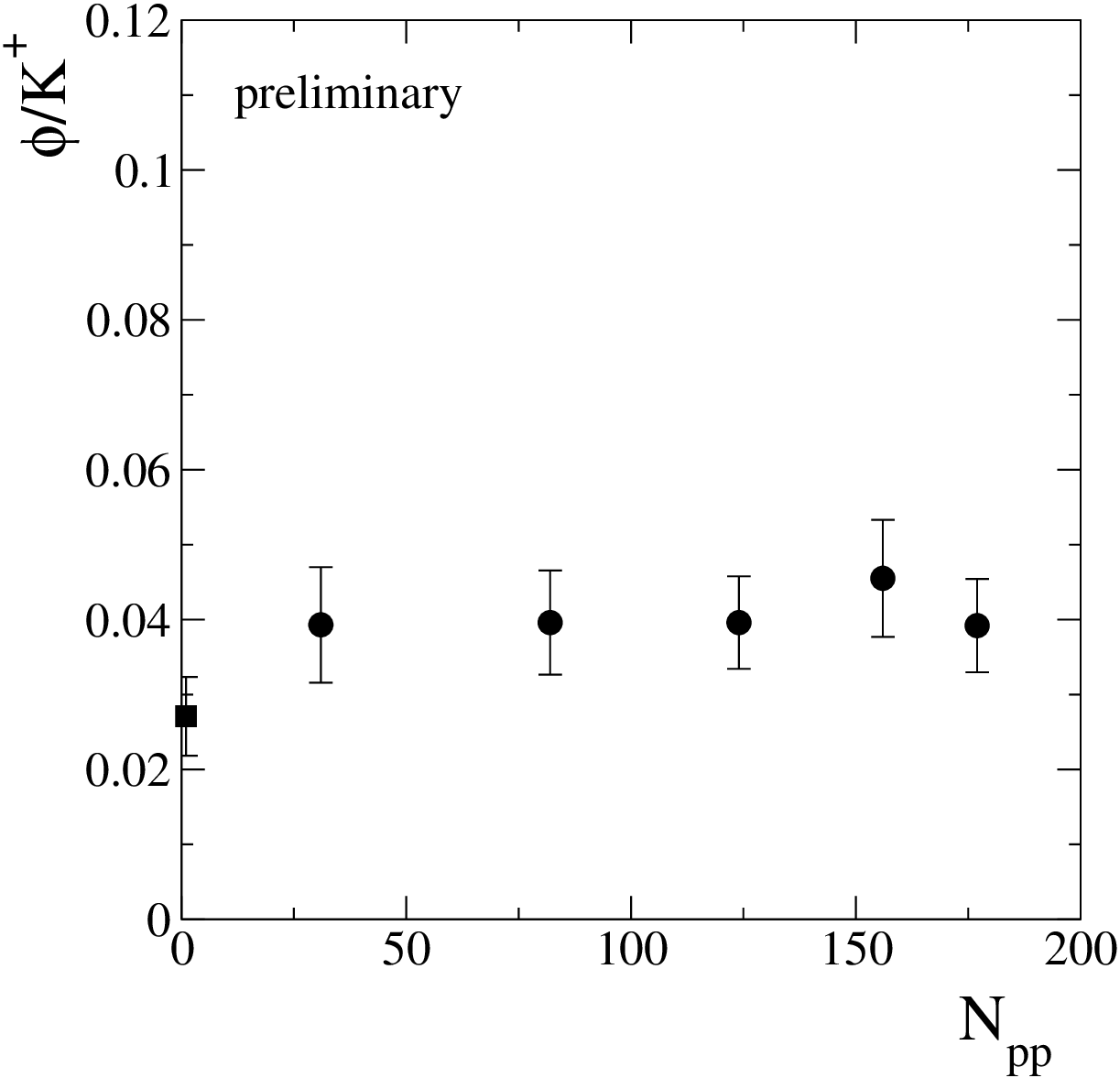}
 \caption{Ratio of $\phi$ to $K^+$ yield as a function of $N_{pp}$.  The
circles are E917 data; the square is p+p from Ref. \cite{phi_pp}.}
 \label{phikplus}
\end{minipage}\hfill
\end{figure}

$\phi$ mesons were reconstructed from $K^+K^-$ pairs in the data 
in a similar fashion to that of
the $\bar{\Lambda}$.
The $\phi$ $m_T$ spectra were fit with 
exponential functions, and the
yields were extracted over the rapidity range $1.2 \leq y \leq 1.6$
for five different centrality classes.  The absolute yields as
a function of centrality have been previously reported~\cite{Wen99}.
Figure~\ref{phinpp} shows the $\phi$ yield divided by
$N_{pp}$, the number of projectile participants, as a function
of $N_{pp}$.  The $\phi/N_{pp}$ ratio increases by a factor of
$2.5$ from peripheral to central collisions, which demonstrates
that $\phi$ mesons are produced through collective phenomena, and
not just a simple superposition of individual nucleon-nucleon 
reactions.  Displayed in Figure~\ref{phikplus} is the
$\phi/K^+$ ratio as a function of $N_{pp}$, where it is evident
that the ratio is independent of centrality.

\section{SUMMARY}

E917 has studied the systematics of baryon stopping, and the production
of $\phi$ mesons, antilambdas, and antiprotons in 
Au+Au collisions at 6, 8, and 10.8 GeV/nucleon.  
These studies indicate that the colliding nuclei are not fully stopped,
and that secondary interactions play a significant role in $\phi$ 
production.  The large increase in the $\bar{\Lambda}$/$\bar{p}$ ratio
from peripheral to central collisions remains unexplained by theory.

\end{document}